\documentclass[12pt]{article}
\usepackage{graphicx}
\usepackage[bookmarksnumbered,colorlinks,plainpages]{hyperref}

\begin{document}

\title{Temperature and pressure in nonextensive thermostatistics}

\author{Q.A. Wang, L. Nivanen, A. Le M\'ehaut\'e, \\
{\it Institut Sup\'erieur des Mat\'eriaux et M\'ecaniques Avanc\'es}, \\
{\it 44, Avenue F.A. Bartholdi, 72000 Le Mans, France} \\
and M. Pezeril \\ {\it Laboratoire de Physique de l'\'etat Condens\'e}, \\
{\it Universit\'e du Maine, 72000 Le Mans, France}}

\date{}

\maketitle

\begin{abstract}
The definitions of the temperature in the nonextensive statistical thermodynamics
based on Tsallis entropy are analyzed. A definition of pressure is proposed for
nonadditive systems by using a nonadditive effective volume. The thermodynamics of
nonadditive photon gas is discussed on this basis. We show that the Stefan-Boltzmann
law can be preserved within nonextensive thermodynamics.
\end{abstract}

{\small PACS : 05.20.-y, 05.70.-a, 02.50.-r}

\section{Introduction}
The nonextensive statistical mechanics (NSM)\cite{Tsal88} based on Tsallis entropy
is believed by many to be a candidate replacing Boltzmann-Gibbs statistics (BGS) for
nonextensive or nonadditive systems which may show probability distributions
different from that of BGS. So according the common belief, NSM, just as BGS, should
be able to address thermodynamic functions and intensive variables like temperature
$T$, pressure $P$, chemical potential $\mu$ etc. Although the Legendre
transformation between the thermodynamic functions is preserved in some versions of
NSM with sometimes certain deformation, the definition of intensive variables is not
obvious if the thermodynamic functions such as entropy $S$, energy $U$ or free
energy $F$ are nonadditive. There are sometimes misleading calculations using
$\beta=1/T=\left(\frac{\partial S}{\partial U}\right)_V$ (let Boltzmann constant
$k=1$) and $P=-\left(\frac{\partial F}{\partial V}\right)_T$ or
$P=\frac{1}{3}\frac{U}{V}$ (for photon gas) without specifying the nonadditivity (or
additivity) of each functions or noticing that additive internal energy $U$ and
volume $V$ associated with nonadditive $S$ and $F$ will lead to non-intensive
temperature or pressure which would make the thermodynamic equilibrium or
stationarity impossible in the conventional sense.

On the other hand, within NSM, due to the fact that different formalisms are
proposed from different statistics or information considerations, thermodynamic
functions do not in general have the same nonadditive nature in different versions
of NSM. This has led to different definitions of, among others, a physical or
measurable temperature $\beta_p$ which is sometimes equal to $\beta$\cite{Wang03c},
sometimes equal to $\beta$ multiplied by a function of the partition function
$Z^{q-1}$\cite{Abe99,Abe01,Mart01,Mart00,Toral} or $Z^{1-q}$\cite{Wang02c,Wang02b}
which keeps $\beta_p$ intensive, where $q$ is the nonadditive entropy
index\footnote{Tsallis entropy is given by $S=\frac{\sum_i p_i^q-1}{1-q}, (q \in
R)$\cite{Tsal88}}, or sometimes defined by deformed entropy and
energy\cite{Wang02b,Wang03b,Wang03z}. This situation often results in confusion and
misleading discussions of these temperatures\cite{Lima03} or other intensive
variables\cite{Nauenberg}, without knowing or mentioning the validity conditions
relevant to them and the risk to have non intensive temperature or pressure.

The present paper tries to make a state of the art on this subject with brief
discussions of the specificities of each formalism of NSM and the relevant
consequences. It is hoped that this paper may offer to the reader a global view of
the situation and of some important questions which are still matters of intense
investigation.

\section{The first definition of physical temperature of NSM}
We look at a composite system containing two subsystems $A$ and $B$, all having the
same $q$ as nonadditive entropy index. The entropy nonadditivity of the total system
is given by
\begin{equation}                                \label{1}
S(A+B)=S(A)+S(B)+(1-q)S(A)S(B).
\end{equation}
This relationship is intrinsically connected with the product joint probability
\begin{equation}                                \label{2}
p_{ij}(A+B)=p_i(A)p_j(B),
\end{equation}
or inversely, where $i$ or $j$ is the index of physical states for $A$ or $B$.
Eq.(\ref{2}) has been intuitively taken as an argument for the independence of $A$
and $B$ and for the energy additivity of $A+B$. This additivity offers the first
possibility to establish zeroth law and to define temperature within
NSM\cite{Abe99,Abe01,Mart01,Mart00,Toral}. The intensive physical temperature is
defined as
\begin{equation}                                        \label{3}
\beta_p=\frac{1}{T_p}=\frac{1}{\sum_{i}^w p_i^q}\frac{\partial{S}}{\partial{U}}
=\frac{1}{\sum_{i}^w p_i^q}\beta.
\end{equation}
This definition is an universal model of NSM and not connected to any specific
statistical formalism.

If this $\beta_p$ is applied to NSM having typically the power law distribution
\begin{equation}                                        \label{4}
p_i=\frac{1}{Z}[1-a\beta_pE_i]^\frac{1}{a}\;\;\;with\;\;[\cdot]\geq 0
\end{equation}
where $E_i$ is the energy of a system at state $i$ and $a$ is $1-q$ or $q-1$
according to the maximum entropy constraints of the formalism\cite{Wang02c,Tsal98},
there may be in general a conflict between the product joint probability and the
energy additivity condition due to the nonadditive energy
$E_i(A+B)=E_i(A)+E_j(B)-a\beta_pE_i(A)E_j(B)$. So the validity of this
thermostatistics strongly lies on neglecting $E_i(A)E_j(B)$.

A mathematical proof\cite{Abe99} shown that this was possible, for a N-body system,
if and only if $q<1$ and $N\rightarrow \infty$. This is not to be forgotten. For the
systems with $q>1$ or with finite size without thermodynamic limits, this additive
energy model is not justified.

Especially, when this model is applied to the formalism of NSM deduced from the
normalized expectation given by the escort probability
$U=\frac{\sum_ip_i^qE_i}{\sum_i p_i^q}$\cite{Tsal98} where $p_i$ is a normalized
probability which reads
\begin{equation}                                        \label{5}
p_i=\frac{1}{Z}[1-(1-q)\beta_p(E_i-U)]^\frac{1}{1-q}
=\frac{1}{Z}[1-(1-q)\frac{\beta}{Z^{1-q}}(E_i-U)]^\frac{1}{1-q},
\end{equation}
Eq.(\ref{3}) becomes
\begin{equation}                                        \label{3x}
\beta_p=\frac{1}{Z^{1-q}}\frac{\partial{S}}{\partial{U}}=Z^{q-1}\beta.
\end{equation}
In this case, $\beta_p$ is not to be confounded with $\beta$ although we have here
$\beta=\frac{\partial S}{\partial U}$ which is evidently non intensive.

\section{The first formalism of NSM}
The first formalism\cite{Tsal88} of NSM maximizes entropy under the constraint
$U=\sum_ip_iE_i$ with normalized $p_i$. The distribution function is given by
\begin{equation}                                        \label{5x}
p_i=\frac{1}{Z}[1-(q-1)\beta_pE_i]^\frac{1}{q-1}.
\end{equation}
The product probability implies the following nonadditivity of energy :
\begin{equation}                                        \label{6}
E_i(A+B)=E_i(A)+E_j(B)-(q-1)\beta_pE_i(A)E_j(B)
\end{equation}
and $U(A+B)=U(A)+U(B)-(q-1)\beta_pU(A)U(B)$. The temperature of this formalism is
still given by Eq.(\ref{3x}) as briefly discussed in \cite{Wang02c}.

The thermodynamic relations can be deduced from the basic expression of entropy of
this formalism
\begin{equation}                                        \label{7}
S=\frac{Z^{1-q}-1}{1-q}+\beta_pZ^{1-q}U
\end{equation}
or $S_p=Z^{q-1}S=\frac{Z^{q-1}-1}{q-1}+\beta_pU$ where $S_p$ is an ``auxiliary
entropy'' introduced to write the generalized heat as $dQ=T_pdS_p$. The first law
reads $dU=T_pdS_p-dW$. The free energy $F$ is defined as
\begin{equation}                                        \label{8}
F=U-T_pS_p=-T_p\frac{Z^{q-1}-1}{q-1}.
\end{equation}
The first law becomes $dF=-S_pdT_p-dW$ where $dW$ is the work done by the system.

$S_p$ can be calculated by using $S_p=-\left(\frac{\partial F}{\partial
T_p}\right)_V$ and Eqs.(\ref{5}) and (\ref{8}) with
$Z=\sum_i[1-(q-1)\beta_pE_i]^\frac{1}{q-1}$\cite{Lima03}. This leads to
\begin{equation}                                        \label{9}
S_p=-\sum_ip_i^{2-q}\frac{p_i^{q-1}-1}{q-1}=\frac{1-\sum_ip_i^{2-q}}{1-q}.
\end{equation}
Notice that this auxiliary entropy is not to be maximized since it is concave only
for $q<2$.

\section{The second formalism of NSM with unnormalized expectation}
This formalism is deduced from the entropy maximum under the constraint
$U=\sum_ip_i^qE_i$ with normalized $p_i$\cite{Curado}. The distribution function is
given by
\begin{equation}                                        \label{5xx}
p_i=\frac{1}{Z}[1-(1-q)\beta_pE_i]^\frac{1}{1-q}.
\end{equation}
and the nonadditivity of energy by $E_i(A+B)=E_i(A)+E_j(B)-(1-q)\beta_pE_i(A)E_j(B)$
and
\begin{equation}                                        \label{5xxxx}
U(A+B)=U(A)Z^{1-q}(B)+U(B)Z^{1-q}(A)+(q-1)\beta_pU(A)U(B).
\end{equation}
As discussed in \cite{Wang03c}, this is the only formalism of NSM in which the
mathematical framework of the thermodynamic relationships is strictly identical to
that of BGS with $\beta_p=\beta$. The heat is given by $dQ=TdS$, the first law by
$dU=TdS-dW$ and the free energy by
\begin{equation}                                        \label{10}
F=U-TS=-T\frac{Z^{1-q}-1}{1-q}.
\end{equation}
Heat and work are interpreted as $dQ=\sum_iE_idp_i^q$ and $dW=\sum_ip_i^qdE_i$,
which is not so simple within other formalisms\cite{Wang03c}.

\section{The formalism with incomplete probability distribution}
If the probability distribution is incomplete in such a way that
$\sum_ip_i^q=1$\cite{Wang01,Wang04} where the sum is calculated only over an
incomplete set of states or of random variables as discussed in \cite{Reny66} and if
we suppose $U=\sum_ip_i^qE_i$, the maximum entropy leads to the following
distribution function
\begin{equation}                                        \label{5xxx}
p_i=\frac{1}{Z}[1-(1-q)\beta_pE_i]^\frac{1}{1-q}.
\end{equation}
where $Z^q=\sum_{i}[1-(1-q)\beta_pE_i]^\frac{1}{1-q}$.

The nonadditivity of energy is given by $$U(A+B)=U(A)+U(B)+(q-1)\beta_pU(A)U(B).$$ The
definition of the physical temperature $\beta_p$ in this formalism is discussed in
\cite{Wang02c,Wang02b} and reads
\begin{equation}                                        \label{3xx}
\beta_p=Z^{1-q}\frac{\partial{S}}{\partial{U}}=Z^{1-q}\beta.
\end{equation}

The introduction of the distribution Eq.(\ref{5xxx}) into Tsallis entropy gives
\begin{equation}                                        \label{7x}
S=\frac{Z^{q-1}-1}{q-1}+\beta_pZ^{q-1}U
\end{equation}
or $S_p=Z^{1-q}S=\frac{Z^{1-q}-1}{1-q}+\beta_pU$ where still $S_p$ is the
``entropy'' introduced to write $dQ=T_pdS_p$. The first law reads $dU=T_pdS_p-dW$
or, with the help of the free energy
\begin{equation}                                        \label{8x}
F=U-T_pS_p=-T_p\frac{Z^{1-q}-1}{1-q},
\end{equation}
$dF=-S_pdT_p-dW$ where $dW$ is the work done by the system. $S_p$ is given
by\cite{Lima03}
\begin{equation}                                        \label{9x}
S_p=-\sum_ip_i^{q}\frac{p_i^{q-1}-1}{q-1}=\frac{1-\sum_ip_i^{2q-1}}{q-1}.
\end{equation}
which is concave only for $q>1/2$ so that not to be maximized to get distribution
functions although its maximum formally leads to $p_i\propto
[1-(q-1)\beta_pE_i]^\frac{1}{q-1}$. Notice that this latter is not the original
distribution function of incomplete statistics.

The above calculation of $S_p$ cannot be carried out for $S$ by using $\beta$ or $T$
because $S\neq -\frac{\partial F}{\partial T}$ although we can write
$F=U-T_pS_p=U-TS$. In addition, $Z$ is not derivable with respect to $\beta$ since
it is a self-referential function when written as a function of $\beta$. This
calculation can be done for $S$ only in the second formalism with unnormalized
expectation and normalized probability associated to
$\beta=1/T=\frac{\partial{S}}{\partial{U}}$.

An additive form of this formalism of a nonadditive statistical thermodynamics is
proposed by using some deformed entropy $s$ and energy $e_i$\cite{Wang02b}, where
$s=\sum_ip_i^q\ln\frac{1}{p_i}$ and $e_i=\frac{\ln[1+(q-1)\beta_p
E_i]}{(q-1)\beta_p}$ both being additive, i.e., $s(A+B)=s(A)+s(B)$ and
$e_{ij}(A+B)=e_i(A)+e_j(B)$. The maximization of $s$ under the constraint
$u=\sum_ip_i^qe_i$ and $\sum_ip_i^q=1$ leads to $p_i=\frac{1}{Z}e^{-\beta_pe_i}$
which is identical to Eq.(\ref{5xxx}). Within this framework, the temperature is
$\beta=\frac{\partial s}{\partial u}$, the deformed first law is $du=T_pds-dw$ ($dw$
is a deformed work), the deformed free energy is
\begin{eqnarray}                                    \label{11x}
f=u-T_ps=-T_p\ln Z=\frac{\ln[1+(q-1)\beta_p F]}{(q-1)\beta_p}.
\end{eqnarray}
In this deformed formalism, everything is just as in BGS. This mathematical
framework has been used for the equilibrium problem of the systems having different
$q$'s\cite{Wang03b,Wang03z}.

\section{Systems having different q's}
The reader should have noticed that all the above discussions are based on the
entropy nonadditivity given by Eq.(\ref{1}) which is valid only for systems having
the same index $q$. For systems $A$, $B$ and $A+B$ each having its own $q$, this
relationship breaks down even if the product joint probability holds. So for
establishing the zeroth law, we need more general nonadditivity for entropy. A
possible one is proposed as follows\cite{Wang03b} :
\begin{eqnarray}                                    \label{1x}
(1-q_{A+B})S(A+B) &=& (1-q_{A})S(A)+(1-q_{B})S(B) \\
\nonumber &+& (1-q_{A})(1-q_{B})S(A)S(B)
\end{eqnarray}
which recovers Eq.(\ref{1}) whenever $q_{A+B}=q_{A}=q_{B}$.

The establishment of zeroth law for this case has been discussed by using the
unnormalized expectations just as in the second formalism of NSM, i.e.,
$u=\sum_ip_i^qe_i$ with $\sum_ip_i=1$\cite{Wang03b}, or $u=\sum_ip_ie_i$ with
$\sum_ip_i^q=1$\cite{Wang03z}. The reason for this is that these unnormalized
expectations allow one to split the thermodynamics of the composite systems into
those of the subsystems through the generalized product joint probability
$p_{ij}^{q_{A+B}}(A+B)=p_i^{q_A}(A)p_i^{q_B}(B)$ if $\sum_ip_i=1$ [or
$p_{ij}(A+B)=p_i(A)p_i(B)$ if $\sum_ip_i^q=1$]. This thermodynamic splitting is just
a necessary condition for the statistical interpretation of the zeroth law.

In this case, the deformed entropy $s$ and energy $u$ are not necessarily additive
as in the case of an unique $q$. In fact, when $u=\sum_ip_i^qe_i$ with $\sum_ip_i=1$
is used, their nonadditivities are given as follows
\begin{equation}                                    \label{1xx}
\frac{q_{A+B}s(A+B)}{\sum_{ij}p_{ij}^{q_{A+B}}(A+B)}=
\frac{q_As(A)}{\sum_{i}p_i^{q_A}(A)}+\frac{q_Bs(B)}{\sum_{j}p_j^{q_B}(B)}
\end{equation}
and
\begin{eqnarray}                                    \label{11}
\frac{q_{A+B}u(A+B)}{\sum_{ij}p_{ij}^{q_{A+B}}(A+B)}=
\frac{q_Au(A)}{\sum_{i}p_i^{q_A}(A)}+\frac{q_Bu(B)}{\sum_{j}p_j^{q_B}(B)}.
\end{eqnarray}
The temperature is given by $\beta_p=\beta=\frac{\partial s}{\partial
u}=\frac{\partial S}{\partial U}$ here $U=\sum_ip_i^qE_i$. The thermodynamic
relations are the same as in the second formalism of NSM or in BGS.

This definition of temperature can be discussed in another way. From Eq.(\ref{1x}),
for a stationary state of $(A+B)$ extremizing $R(A+B)$, we have
\begin{eqnarray}                                    \label{1xxa}
\frac{(q_{A}-1)dS(A)}{\sum_ip_i(A)}+\frac{(q_{B}-1)dS(B)}{\sum_ip_i(B)}=0.
\end{eqnarray}
Now using the above mentioned product joint probability and the relationship
$\sum_ip_i^q=Z^{1-q}+(1-q)\beta U$, we get
$\frac{(1-q_{A})\beta(A)dU(A)}{\sum_ip_i(A)}
+\frac{(1-q_{B})\beta(B)dU(B)}{\sum_ip_i(B)}=0$ which suggests following energy
nonadditivity
\begin{eqnarray}                                    \label{14xx}
\frac{(1-q_{A})dU(A)}{\sum_ip_i(A)}+\frac{(1-q_{B})dU(B)}{\sum_ip_i(B)}=0
\end{eqnarray}
as the analogue of the additive energy $dU(A)+dU(B)=0$ of Boltzmann-Gibbs
thermodynamics. Eq.(\ref{13}) and Eq.(\ref{14}) lead to $\beta(A)=\beta(B)$.

Summarizing the definitions of temperature, we have
$\beta_p=Z^{q-1}\beta=Z^{q-1}\frac{\partial S}{\partial U}$ for the normalized
expectations $U=\sum_ip_iE_i$ or $U=\sum_ip_i^qE_i/\sum_ip_i^q$ with $\sum_ip_i=1$,
and $\beta_p=Z^{1-q}\beta=Z^{1-q}\frac{\partial S}{\partial U}$ for the normalized
expectations $U=\sum_ip_i^qE_i$ with $\sum_ip_i^q=1$. On the other hand,
$\beta_p=\beta=\frac{\partial S}{\partial U}$ can be preserved if and only if
unnormalized expectations $U=\sum_ip_i^qE_i$ with $\sum_ip_i=1$ (or $U=\sum_ip_iE_i$
with $\sum_ip_i^q=1$) are used. The additive energy model of the nonextensive
thermostatistics is justified for $q<1$ and with the thermodynamic limits.

\section{What about the pressure?}
If the work in the first law is $dW=PdV$, where $P$ is the pressure and $V$ a
certain volume, then the pressure can be calculated through
$P=-\left(\frac{\partial{F}}{\partial{V}}\right)_T$. If we want the pressure to be
intensive, $V$ will be nonadditive. This is a delicate choice to make since
nonadditive volume is nontrivial and not so easy to be understood as nonadditive
energy or entropy. For a standard system, we tend to suppose additive volume as well
as additive particle number. However, in view of the fact that the work $dW$ is in
general nonadditive, additive volume implies non intensive pressure $P$, which is
impossible if the equilibrium or stationary state is established in the conventional
sense for, e.g. a gas of photons or of other particles. So, first of all, for the
following discussion, let us suppose an intensive pressure $P$, i.e., $P(A)=P(B)$ at
equilibrium or stationarity.

Intensive $P$ implies nonadditive $V$. If one wants to suppose additive volume (the
real one) and particle number $N$, $V$ must be regarded as an effective volume, as a
function of the real volume $V_p$ supposed additive.

In this case, a question arises about the nature of the work $dW$ which is no more
proportional to the real volume $dV_p$. Is it a real work? Our answer is Yes because
$dW$ is supposed to contribute to the energy variation $dU$ or $dF$ according to the
first law. A possibility to account for this work is that, for a nonextensive or
nonadditive system, e.g., a small system or a heterogeneous system, the
surface/interface effects on the total energy, compared with the volume effect, are
not negligible. When the pressure makes a small volume variation $dV_p$, the work
may be $dW=PdV_p+dW_\sigma$ where $dW_\sigma$ is the part of work related to the
surface/interface variation $d\sigma$. In general, the relationship $dW_\sigma\sim
d\sigma$ should depend on the nature and the geometry of the system of interest. If
we suppose a simple case where $dW_\sigma=\alpha Pd(\sigma^\theta)$ and
$\sigma=\gamma V_p^\eta$ ($\alpha$, $\gamma$, $\eta$ and $\theta$ are certain
constants), the work can be written as $dW=PdV_p+\alpha\gamma
Pd(V_p^{\eta\theta})=Pd[V_p+\alpha\gamma V_p^{\eta\theta}]$ which means
$V=V_p+\alpha\gamma V_p^{\eta\theta}$. This example shows that a nonadditive
effective volume can be used for nonextensive systems to write the nonadditive work
in the form $dW=PdV$, just as in the conventional additive thermodynamics.

\subsection{A definition of pressure for NSM}
Now let us come back to NSM. To determine the nonadditivity of the effective volume
$V$ with additive real volume $V_p$, one has to choose a given version of NSM with
given nonadditivity of entropy and energy. Without lose of generality, the following
discussion will be made within the second formalism of NSM. From the entropy
definition and nonadditivity Eq.(\ref{1}) and the energy nonadditivity
Eq.(\ref{5xxxx}), we can write, at equilibrium or stationarity,

\begin{eqnarray}                                        \label{12}
dS(A+B) & = & [1+(1-q)S(B)]dS(A)+ [1+(1-q)S(A)]dS(B) \\ \nonumber
    & = & \sum_{i}p_i^q(B)\left[
\left(\frac{\partial S(A)}{\partial U(A)}\right)_V dU(A) +\left(\frac{\partial
S(A)}{\partial V(A)}\right)_UdV(A) \right] \\ \nonumber &+&
\sum_{i}p_i^q(A)\left[\left(\frac{\partial S(B)}{\partial U(B)}\right)_V dU(B)
+\left(\frac{\partial S(B)}{\partial V(B)}\right)_UdV(B)\right] \\ \nonumber
&=&\sum_{i}p_i^q(B)\left[\left(\frac{\partial S(A)}{\partial U(A)}\right)_V
-\left(\frac{\partial S(B)}{\partial U(B)}\right)_V\right]dU(B) \\ \nonumber
&+&\sum_{i}p_i^q(B)\left(\frac{\partial S(A)}{\partial U(A)}\right)_V
\left(\frac{\partial U(A)}{\partial V(A)}\right)_SdV(A) \\\nonumber
&+&\sum_{i}p_i^q(A)\left(\frac{\partial S(B)}{\partial U(B)}\right)_V
\left(\frac{\partial U(B)}{\partial V(B)}\right)_SdV(B) \\\nonumber
&=&\beta\left[P(A)\sum_{i}p_i^q(B)dV(A)+P(B)\sum_{i}p_i^q(A)dV(B)\right]=0.
\end{eqnarray}
Here we have used $\frac{dS(A)}{\sum_{i}p_i^q(A)}+\frac{dS(B)}{\sum_{i}p_i^q(B)}=0$,
$\frac{dU(A)}{\sum_{i}p_i^q(A)}+\frac{dU(B)}{\sum_{i}p_i^q(B)}=0$\cite{Wang03c}, and
$\left(\frac{\partial S}{\partial V}\right)_U=\left(\frac{\partial U}{\partial
V}\right)_S\left(\frac{\partial S}{\partial U}\right)_V$. Then $P(A)=P(B)$ leads to
$\frac{dV(A)}{\sum_{i}p_i^q(A)}+\frac{dV(B)}{\sum_{i}p_i^q(B)}=0$, which implies
that the quantity $\frac{dV}{\sum_{i}p_i^q}$ is additive, just as
$\frac{dS}{\sum_{i}p_i^q}$ and $\frac{dU}{\sum_{i}p_i^q}$.

It can be checked that this kind of calculation is also possible within other
versions of NSM as long as the energy nonadditivity is determined by the product
joint probability which is in turn a consequence of the entropy nonadditivity
Eq.(\ref{1}) or Eq.(\ref{1x}) postulated for Tsallis entropy.

\subsection{About nonadditive photon gas}
Now let us suppose a nonadditive photon gas, which is possible when emission body is
small. For example, the emission of nanoparticles or of small optical cavity whose
surface/interface effect may be important. We have seen in the above paragraph that
$dU$, $dS$ and $dV$ should be proportional to each other. This can be satisfied by
$U=f(T)V$ and $S=g(T)V$. In addition, we admit the photon pressure given by
$P=\frac{U}{3V}=\frac{1}{3}f(T)$. From the first law $dU=TdS-PdV$, we obtain
\begin{equation}                                        \label{13}
V\frac{\partial f}{\partial T}dT+fdV=T(V\frac{\partial g}{\partial
T}dT+gdV)-\frac{1}{3}fdV,
\end{equation}
which means $\frac{\partial f}{\partial T}=T\frac{\partial g}{\partial T}$ and
$\frac{4}{3}f=Tg$ leading to $\frac{1}{3}\frac{\partial f}{\partial
T}=\frac{4f}{3T}$ implying
\begin{equation}\label{14}
f(T)=cT^4
\end{equation}
where $c$ is a constant. This is the Stefan-Boltzmann law. On the other hand, from
the relationship $(\frac{\partial S}{\partial V})_T=(\frac{\partial P}{\partial
T})_V$, we obtain $g=\frac{1}{3}\frac{\partial f}{\partial T}$ and $g(T)=bT^3$ where
$b$ is a constant. Notice that the above calculation is similar to that in the
conventional thermodynamics. This is because the thermodynamic functions here,
though nonadditive, are nevertheless ``extensive'' with respect to the effective
volume. This result contradicts what has been claimed for blackbody radiation on the
basis of non intensive pressure\cite{Nauenberg}, and is valid as far as the pressure
is intensive.

Is intensive pressure always true? The final answer of course depends on
experimental proofs which are still missing as far as we know. If pressure may be
non intensive for nonadditive or nonextensive systems, the whole theory of
thermodynamics must be reviewed.

\section{Conclusion}
In summery, we have analyzed all the temperature definitions of NSM we can actually
find in the literature. A definition of intensive pressure is proposed for
nonextensive thermodynamics by using a nonadditive effective volume. The
thermodynamics of a nonadditive photon gas is discussed on that basis. It is shown
from purely thermodynamic point of view that the Stefan-Boltzmann law can be valid
within NSM in this case.

\end{document}